\documentstyle[11pt,newpasp,twoside]{article}
\def\edcomment#1{\iffalse\marginpar{\raggedright\sl#1\/}\else\relax\fi}
\reversemarginpar

\begin{document}
\title{Temperature Variations and Abundance Determinations
in Planetary Nebulae}
\author{Silvia Torres-Peimbert and Manuel Peimbert} \affil{Instituto de
Astronom\'{\i}a, Universidad Nacional Aut\'onoma de M\'exico; Apdo. postal
70--264; Ciudad Universitaria; M\'exico D.F. 04510; M\'exico.}

\begin{abstract}
It is argued that an important fraction of PNe present large temperature
variations that are not due to observational errors nor to incomplete atomic
physics. Seven possible causes for these variations are reviewed, one of them
is presented for the first time.
\end{abstract}

\section{Overview}

Photoionization models for chemically homogeneous gaseous nebulae of constant
density predict an almost constant temperature and consequently very often
observers assume a constant temperature to determine chemical abundances. There
is growing observational evidence that indicates the presence of large
temperature variations in gaseous nebulae, in contradiction with the models
mentioned above.  Consequently the relevance of temperature variations in the
abundance determinations has to be considered in detail.

In section 2 we review the observational evidence in favor of large temperature
variations. In section 3 we review different possible explanations for the
observed variations. In section 4 we discuss the implications of the
temperature variations in the abundance determinations, these implications
depend on the source of the temperature variations.

Recent reviews on the temperature structure of gaseous nebulae are those of
Peimbert, M. (1995, 2002), Esteban (2002), Stasi\'nska (2002), and Liu (2002a,
2002b).

\section{Temperature Variations}

\subsection{Definitions}

The average temperature, $T_0$, and the mean square temperature fluctuation,
$t^2$, are given by
\begin{equation}
T_0 (N_e, N_i) = \frac{\int T_e({\bf r}) N_e({\bf r}) N_i({\bf r}) dV}
{\int N_e({\bf r}) N_i({\bf r}) dV},
\end{equation}
and
\begin{equation}
t^2 = \frac{\int (T_e - T_0)^2 N_e N_i dV}{T_0^2 \int N_e N_i dV},
\end{equation}
respectively, where $N_e$ and $N_i$ are the electron and the ion densities of
the observed emission line and $V$ is the observed volume (Peimbert 1967).

To determine $T_0$ and $t^2$ we need two different methods to derive $T_e$: one
that weighs preferentially the high temperature regions and one that weighs
preferentially the low temperature regions. For example the temperature derived
from the ratio of the [O~III] $\lambda\lambda$ 4363, 5007 lines,
$T_{(4363/5007)}$, and the temperature derived from the ratio of the Balmer
continuum to $I(H\beta)$, $T_{({\rm Bac}/H\beta)}$, that are given by
\begin{equation}
T_{(4363/5007)} = T_0 \left[ 1 + {\frac{1}{2}}\left({\frac{90800}{T_0}} - 3
\right) t^2\right],
\end{equation}
and
\begin{equation}
T_{({\rm Bac}/{\rm H}\beta)} = T({\rm Bac}) = T_0 (1 - 1.70 t^2),
\end{equation}
respectively. It is also possible to use the intensity ratio of a collisionally
excited line of an element $p+1$ times ionized to a recombination line of the
same element $p$ times ionized, this ratio is independent of the element
abundance and in the low density limit depends only on the electron
temperature. Two examples of this method are:
\begin{equation}
T_{({\rm O~II}rec/{\rm O~III}coll)} = T_{(4649/5007)} = f_1(T_0,t^2)
\end{equation}
and
\begin{equation}
T_{({\rm C~II}rec/{\rm C~III}coll)} = T_{(4267/1909)} = f_2(T_0,t^2),
\end{equation}
by combining these ratios with a temperature determined from the ratio of two
collisionally excited lines, like $T_{(4363/5007)}$, it is also possible to
derive $T_0$ and $t^2$.

From the ratio of two He~I recombination lines it is possible to obtain
relationships of the type:
\begin{equation}
T_{({\rm He~I}, i/j)} = f_{ij}(T_0,t^2,N_e,\tau_{3889}),
\end{equation}
and from a set of 5 or more helium lines (which should include $\lambda\lambda$
3889, 6678, and 7065) it is possible to obtain $T_0$, $t^2$, $N_e$, and
$\tau_{3889}$ (see Peimbert, Peimbert, \& Ruiz 2000). Sometimes the errors in
these determinations are quite large, in such case it is possible to combine
the information from the helium lines with the one from $T_{(4363/5007)}$ to
obtain better determinations of $T_0$ and $t^2$ (see Peimbert, Peimbert, \&
Luridiana 2002).

\subsection{Errors in the temperature determinations}

Different types of errors are present in the determinations of the electron
temperatures, and in some cases these errors have led to erroneous temperature
determinations. Before accepting that temperature variations are present in a
given object, it is necessary to rule out any possible sources of error.

Rola \& Stasi\'nska (1994) and Rola \& Pelat (1994) have argued that there is a
systematic overestimation of the line intensities of lines with low signal to
noise ratios that leads to the overestimation of the $T_{(4267/1909)}$
temperatures. This effect is present in many objects and has to be taken into
consideration, but for many well observed objects it is not the main cause of
the derived large $T_{(4363/5007)}$ -- $T_{(4267/1909)}$ differences
(e.g., Peimbert, Torres-Peimbert, \& Luridiana 1995).

The discrepancy between the C abundances derived from the C~II $\lambda$ 4267
recombination line and those derived from the collisionally excited 1906+1909
C~III lines has been known for at least 20 years and the idea that the large
intensity of the C~II $\lambda$ 4267 line was due to an incomplete knowledge of
the relevant atomic physics has been mentioned on several
occasions. Nevertheless similar excesses in the intensities of recombination
lines are observed in different multiplets of the same ion and in other
elements (Liu et al. 1995; Mathis, Torres-Peimbert, \& Peimbert 1998).

Viegas \& Clegg (1994) suggested that the differences between $T_{(4649/5007)}$
and $T$(Bac) could be due to density fluctuations diminishing the $I(4959)$ and
$I(5007)$ values by collisional de-excitation yielding a spuriously high
$T_{(4649/5007)}$ value. This is indeed the case for the planetary nebula
M2-29, where from ground based observations of the whole object a
$T_{(4363/5007)}$ of 24,000~K and a $N_e = 2000$~cm$^{-3}$ are derived, while
from HST observations it is found that it has at least two density components
(one with $N_e \sim 10^4$ cm$^{-3}$ and another with $N_e \sim 10^6$~cm$^{-3}$,
the $T_e$ in the low density component is of 8800~K); the high temperature
derived from the ground observations is spurious and yields O/H$ = 5.1 \times
10^{-5}$, a factor of ten smaller than the value derived adopting a two density
component model (Pe\~na, Torres-Peimbert, \& Ruiz 1991; Torres-Peimbert et
al. 2002). A similar result has been found for Mz~3 by Zhang \& Liu (2002).

\subsection{Observations of Temperature Variations}

From uniform density chemically homogeneous photoionization models it has been
found that $t^2$ is in the 0.000 to 0.025 range, with typical values around
0.005 (e.g., Gruenwald \& Viegas 1995).

From observations and by combining equation 3 with any of the equations 4 to 7
it is possible to determine $t^2$ and $T_0$. These combinations indicate values
of $t^2$ in the 0.00 to 0.15 range with typical values around 0.04.  The errors
in the best $t^2$ determinations given in the literature are typically in the
0.01 to 0.02 range.

Some of the best determinations of $t^2$ and $T_0$ are: from equations 3 and 4
those by Peimbert (1971), Liu \& Danziger (1993), Liu et al. (1995, 2000,
2001); from equations 3 and 5 those by Peimbert, Storey, \& Torres-Peimbert
(1993), Liu et al. (1995, 2000, 2001); from equations 3 and 6 those by Rola \&
Stasi\'nska (1994), Peimbert, et al. (1995); and from equations 3 and 7 those
by Peimbert et al. (1995, 2000, 2002).

Liu et al.(2001, see also Peimbert, A. 2002), presented a good correlation
between $Log({\rm O}^{++}/{\rm H}^+)_{RL}/({\rm O}^{++}/{\rm H}^+)_{CEL}$ and
$T_{(4363/5007)}$--$T$(Bac) and mention that this correlation strongly supports
the idea that the temperature variations are real. In Table 1 we present the
$T_{(4363/5007)}$ and $T({\rm Bac})$ temperatures derived by Liu et al. (2000,
2001) for three of the objects that show very large temperature
variations. From these values we have computed $T_{(4649/5007)}$ and from
equations 3 and 4 we have derived $t^2$(3,4), and from equations 3 and 5 we have
derived $t^2$(3,5).  From this table it can be seen that $t^2$(3,4) and
$t^2$(3,5) are not only proportional to each other but that they are in good
agreement with each other. This implies that most of the temperature
differences among $T({\rm Bac})$, $T_{(4363/5007)}$, and $T_{(4649/5007)}$ are
due to real temperature variations in the nebulae, thus an explanation for the
origin of these variations has to be sought.

\begin{table}
\begin{center}
Table 1 \\
Mean Square Temperature Fluctuation Values\\
\begin{tabular}{lccccc}
\\
\hline
\hline
\rule[-3mm]{0mm}{8mm}
Object & $T_{(4363/5007)}$ & $T_{({\rm Bac}/{\rm H}\beta)}$ &
$T_{(4649/5007)}$ & $t^2(3,4)$ & $t^2(3,5)$ \\ [0.2ex]
\hline
\\
M1-42    & 9220 & 3560 & 5550 & 0.123 & 0.120 \\
M2-36    & 8380 & 5900 & 6025 & 0.053 & 0.078 \\
NGC 6153 & 9140 & 6080 & 6220 & 0.065 & 0.097 \\
\\
\hline
\hline
\end{tabular}
\end{center}
\end{table}

\section{What Causes the Temperature Variations?}

In what follows we present seven different mechanisms that can produce large
temperature variations. It is clear that all of them should be
considered. Their relative importance probably varies for each object,
and in many cases, the ones that show $t^2 < 0.01$ values, all of them are
negligible.

\subsection{Shadowed Regions}

To explain the intensities of lines of low degree of ionization Mathis (1976)
suggested the existence of regions shadowed from the central stars, these
regions would be ionized by diffuse radiation and would have a considerably
smaller temperature than those regions ionized directly from the central stars
of H~II regions.

\subsection{Chemical Inhomogeneities}

Some PNe, like A30 and A78, present important chemical inhomogeneities;
therefore the idea that temperature variations can be due to various degrees of
chemical inhomogeneities has been proposed. Chemically inhomogeneous
photoionization models have been presented by different authors (e.g.,
Torres-Peimbert, Peimbert, \& Pe\~na 1990; Kingdon \& Ferland 1998; Pequignot
et al. 2002a, 2002b; Tylenda 2002).

\subsection{Density Variations}

Extreme density variations are present in most PNe, as can be seen from optical
images; these clumps, shells, and filaments, can produce temperature variations
by themselves under the assumption of photoionization heating only. Mihalszky
\& Ferland (1983) have presented chemically homogeneous photoionization models,
including density variations in the $10^2$ to $10^{4.3} {\rm cm}^{-3}$ range,
and find values of $t^2$ of about 0.005, which could explain only a small
fraction of the largest discrepancies observed.
 
\subsection{Deposition of Mechanical Energy}

Shocks have been proposed as a mechanism to account for temperature
fluctuations (Peimbert, Sarmiento, \& Fierro 1991; Peimbert et al. 1995 and
references therein). Peimbert et al. (1995) have found that those objects that
show higher velocity dispersions are the ones that show higher temperature
variations, supporting the presence of shocks.  In a PN a shock front is
expected to be formed at the interface between a fast stellar wind from the
central star and the slowly expanding circumstellar envelope. The fast stellar
wind would carve out a hot bubble in the interior of the nebula that would emit
primarily in X-rays. Clear evidence for such hot bubbles has been found
recently for a small number of PNe from X-ray imaging (Kastner et al. 2000;
Kastner, Vrtilek, \& Soker 2001; Chu et al. 2001).

Mellema (1997) finds strong temperature variations from hydrodynamical
simulations of bipolar PNe. The dense gas near the equator has temperatures
around 9000~K while the gas near the poles has temperatures between 20,000 and
50,000~K. The reasons are two: at the poles the nebula has a higher shock
velocity, leading to more heating; and a lower density, leading to less
efficient cooling. Bohigas (2002) finds that in the outer regions of 9 PNe of
Type I the $T({\rm O}^{++})/T({\rm N}^+)$ is large and the H$\alpha$/[S~II]
line ratio is small, a typical combination in shock excited plasmas.

\subsection{Deposition of Magnetic Energy}

Ferland, in the meeting on {\it Ionized Gaseous Nebulae} (Mexico City 2000),
suggested that magnetic reconnection can provide localized temperature
variations in ionized plasmas. According to Garc\'{\i}a-Segura, L\'opez, \&
Franco (2001) the multiple, regularly spaced concentric shells around some PNe
could be due to the effects a of a solar-like magnetic cycle, with periodic
polarity inversion, in the slow wind of an asymptotic giant branch (AGB)
star. Presumably these shells of alternating polarity could give rise to
magnetic reconnection processes once that they are compressed in the formed PN
(i.e. in the swept-up shell).  Magnetic fields of the order of milligauss have
been measured in the torus surrounding K3-35 (Miranda et al. 2002); also the
presence of a magnetic field has been inferred in the planetary nebula
OH0.9+1.3 (Zijlstra et al. 1989) and in the pre-planetary nebula
IRAS17150$-$3224 (Hu et al. 1993).

\subsection{Dust Heating}

Stasi\'nska \& Szczerba (2001) have analyzed the effects of photoelectric
heating by dust grains in photoionization models of PNe. They have shown that
this process is particularly important if filamentary PNe contain a population
of small grains. The temperature structure of such dusty and filamentary
nebulae would produce large temperature variations.

\subsection{Decrease of the Ionizing Flux}

The ionizing flux from the central stars of PNe is not constant.  As a first
approximation the ionization increases, then reaches a maximum and finally
decreases. During the decrease and in those directions where the nebula is
density bounded there would be ionized regions that would become isolated from
the stellar ionizing photons, these regions would cool first and would
recombine later (e.g., Spitzer 1978). A typical recombination time is $\sim
10^5/N_e$ years while the cooling time is an order of magnitude smaller,
therefore cool regions with a relatively high degree of ionization are expected
to be present.

This effect might be important for PNe produced by massive stars, because the
central stars evolve relatively fast. One possible example of this mechanism
might be provided by N66 in the LMC. In this object the ionizing flux has
diminished by a factor of two in the last ten years (Pe\~na et al. 1997;
Pe\~na, Hamann, \& Ruiz 2002) and the RL/CEL abundance ratio is of about an
order of magnitude (Tsamis et al. 2002). Other PNe central stars have presented
important flux changes in timescales of a few years (e.g., Kostiakova 2002 and
references therein). It is possible that this mechanism might account for the
very low values of $T$(Bac), $\sim$3000~K, found by Luo \& Liu (2002) in the
outer parts of NGC 7009.

It would be possible also to isolate regions from the stellar ionizing photons
if the amount of material in the line of sight increases with time, even if the
stellar ionizing radiation remains constant. It has been suggested that dust
clouds moving tangentially to the line of sight are responsible for the
1981--1985 and 1996--1997 fadings of V651 Mon, the central star of NGC 2346
(Costero et al. 1986; Kato, Nogami, \& Baba 2001). In addition to dust clouds a
similar effect would be caused by tangentially moving gaseous clouds of higher
density than the ambient density.

\section{Discussion}

Temperature variations higher than those predicted by chemically homogeneous
photoionization models of constant density are definitively present in a good
fraction of PNe.

Chemically homogeneous photoionization models of constant density provide a
good first approximation to the temperature structure and to the chemical
abundances only for about half of the well observed nebulae. However, for the
other half, it is paramount to have observations of high accuracy to determine
their temperature structure which is needed to obtain accurate abundances of
heavy elements relative to hydrogen.

Spatially resolved values of $T_0$ and $t^2$ are needed to decide among the
various possibilities mentioned above. For example, if $T_0$ decreases in the
outer regions, the temperature variations might be due to a decrease of the
ionizing flux; in the opposite case the temperature variations could be due to
shock heating.

A very important observational clue to study this problem is provided by
Garnett \& Dinerstein (2001) who find a large range in the difference between
the O$^{++}$ abundances derived from O~II and those derived from [O~III],
spanning from no difference up to a factor of six. The size of this discrepancy
is anti-correlated with nebular surface brightness; compact,
high-surface-brightness nebulae have the smallest discrepancies.

It is important to determine the cause of the temperature variations to be able
to constrain the models of stellar evolution and of galactic chemical
evolution. If the temperature fluctuations due to chemical inhomogeneities
dominate, the proper abundances to use are those derived from forbidden lines
under the assumption that $t^2 \sim 0.01$; alternatively if the temperature
fluctuations due to the combination of the other six mechanisms mentioned above
dominate, the proper abundances to use are those provided by the recombination
lines.

The C/H values derived from recombination lines in PNe are in better agreement
with chemical evolution models of the Galaxy than the values derived from
collisionally excited lines under the assumption that $t^2 = 0.000$ (Carigi
2002). Alternatively the abundances derived from the infrared lines seem to be
in favor of chemical inhomogeneities, but these abundances could be higher if
large density fluctuations are present.

The presence of abundance gradients in the Galaxy and the values of $\Delta
Y/\Delta Z$ and $\Delta Y/\Delta O$ derived from galactic and extragalactic
H~II regions indicate that there are significant temperature variations in H~II
regions and that they are not due to chemical inhomogeneities (see Peimbert \&
Peimbert 2002 for a review).

\acknowledgements{We are grateful to G. Garc\'{\i}a-Segura, A. Peimbert, and
M. Pe\~na, for several excellent suggestions, and to our Australian colleagues
for an outstanding meeting.}


\begin{references}

\reference{Bohigas, J. 2002, these proceedings}

\reference{Carigi, L. 2002, in Ionized Gaseous Nebulae, RevMexAACS, 12,
234}

\reference{Chu, Y.-H., Guerrero, M. A., Gruendl, R. A., Williams, R. M., \&
Kaler, J. B. 2001, \apj, 553, L69}

\reference{Costero, R., Tapia, M., Echevarr\'{\i}a, J., Roth, M., Quintero, A.,
\& Barral, J. F. 1986, RMexAA, 13, 149}

\reference{Esteban, C. 2002, in Ionized Gaseous Nebulae, RevMexAACS, 12, 56}

\reference{Garc\'{\i}a-Segura, G., L\'opez, J. A., \& Franco, J. 2001, \apj,
560, 928}

\reference{Garnett, D. R., \& Dinerstein, H. L. 2001, RevMexAASC, 10, 13}

\reference{Gruenwald, R., \& Viegas, S.M. 1995, \aap, 303, 535}

\reference{Hu, J. Y., Slijkhuis, S., Rieu, N.-Q., \& de Jong, T. 1993, \aap,
273, 185 }

\reference{Kastner, J. H., Soker, N., Vrtilek, S. D., \& Dgani, R. 2000, \apj,
545, L57}

\reference{Kastner, J. H., Vrtilek, S. D., \& Soker, N. 2001, \apj, 550, L189}

\reference{Kato, T., Nogami, D., \& Baba, H. 2001, \pasj, 53, 901}

\reference{Kingdon, J. B., \& Ferland, G. J. 1998, \apj, 506, 323}

\reference{Kostyakova, E. B. 2002, in Ionized Gaseous Nebulae, RevMexAACS,
12, 167}

\reference{Liu, X.-W. 2002a, in Ionized Gaseous Nebulae, RevMexAACS, 12,
70}

\reference{Liu, X.-W. 2002b, these proceedings}

\reference{Liu, X.-W., \& Danziger, I. J. 1993, \mnras, 263, 256}

\reference{Liu, X.-W., Luo, S.-G., Barlow, M. J., Danziger, I. J., \& Storey,
P. J. 2001, \mnras, 327, 141}

\reference{Liu, X.-W., Storey, P. J., Barlow, M. J., \& Clegg, R. E. S. 1995,
\mnras, 272, 369}

\reference{Liu, X.-W., Storey, P. J., Barlow, M. J., Danziger, I. J.,
Cohen, M., \& Bryce, M. 2000, \mnras, 312, 585}

\reference{Luo, S.-G., \& Liu, X.-W 2002, these proceedings}

\reference{Mathis, J. S. 1976, \apj, 207, 442}

\reference{Mathis, J. S., Torres-Peimbert, S., \& Peimbert, M. 1998, \apj, 495,
328}

\reference{Mellema, G. 1997, in IAU Symposium 180, eds. H. J. Habing \&
H. J. G. L. M. Lamers, p. 262}

\reference{Mihalszky, J. S., \& Ferland, J. 1983, \pasp, 95, 284}

\reference{Miranda, L. F., G\'omez, Y., Anglada, G., \& Torrelles, J. M. 2001,
Nature, 414, 284 }

\reference{Peimbert, A. 2002, in preparation}

\reference{Peimbert, A., Peimbert, M., \& Luridiana, V. 2002, \apj, 565, 668}

\reference{Peimbert, M. 1967, \apj, 150, 825}
 
\reference{Peimbert, M. 1971, Bol. Obs. Tonantzintla y Tacubaya, 6, 29}
 
\reference{Peimbert, M. 1995, in The Analysis of Emission Lines, eds.
R. E. Williams, \& M. Livio (Cambridge), 165}
 
\reference{Peimbert, M. 2002, in Ionized Gaseous Nebulae, RevMexAACS, 
12, 275}
 
\reference{Peimbert, M., \& Peimbert, A. 2002, RevMexAASC, in press}
 
\reference{Peimbert, M., Peimbert, A., \& Ruiz, M. T. 2000, \apj, 541, 688}

 
\reference{Peimbert, M., Sarmiento , A., \& Fierro, J. 1991, \pasp, 103, 815}
 
\reference{Peimbert, M., Storey , P. J., \& Torres-Peimbert, S. 1993, \apj,
414, 626}
 
\reference{Peimbert, M., Torres-Peimbert, S., \& Luridiana, V. 1995, RevMexAA,
31, 131}
 
\reference{Pe\~na, M., Hamann, W.-R., Koesterke, L., Maza, J., M\'endez, R. H.,
Peimbert, M., Ruiz, M. T., \& Torres-Peimbert, S.  1997, \apj, 491, 233}

\reference{Pe\~na, M., Hamann, W.-R., \& Ruiz, M. T. 2002, these proceedings}

\reference{Pe\~na, M., Torres-Peimbert, S., \& Ruiz, M. T. 1991, \pasp, 103,
865}
 
\reference{P\'equignot, D., Amara, M., Liu, X.-W., Barlow, M. J., Storey, P. J.,
Morisset, C., Torres-Peimbert, S., \& Peimbert, M. 2002a, in Ionized Gaseous
Nebulae, RMexAASC, 12, 142}

\reference{P\'equignot, D., Liu, X.-W., Barlow, M. J., Storey P. J., \& Morisset,
C. 2002b, these proceedings}
 
\reference{Rola, C., \& Pelat, D. 1994, \aap, 287, 676}

\reference{Rola, C., \& Stasi\'nska, G. 1994, \aap, 282, 199}
 
\reference{Spitzer L. Jr. 1978, in Physical Processes in the 
Interstellar Medium (New York: Wiley-Interscience), p. 139}

\reference{Stasi\'nska, G. 2002, in Ionized Gaseous Nebulae, RevMexAACS, 12, 62}
 
\reference{Stasi\'nska, G., \& Szczerba, R. 2001, \aap, 379, 1024}
 
\reference{Torres-Peimbert, S., Dufour, R. J., Peimbert, M., \& Pe\~na,
M. 2002, in preparation}
 
\reference{Torres-Peimbert, S., Peimbert, M., \& Pe\~na, M.: 1990, \aap, 233,
540}
 
\reference{Tsamis, Y. G., Barlow, M. J., Liu, X.-W., \& Danziger, I. J. 2002,
these proceedings}

\reference{Tylenda, R. 2002, these proceedings}

\reference{Viegas, S. M., \& Clegg, R. E. S. 1994, \mnras, 271, 993}
 
\reference{Zhang, Y., \& Liu, X.-W. 2002, these proceedings}
 
\reference{Zijlstra, A. A., Te Lintel Hekkert, P., Pottasch, S. R.,
 Caswell, J. L., Ratag, M., \& Habing, H. J. 1989 \aap, 217, 157}
 
\end{references}
\end{document}